\documentclass[epsf,amssymb,aps,prd,nofootinbib,floatfix,superscriptaddress,secnumarabic]{revtex4}

\usepackage{bm}% bold math                                                                              
\usepackage{amsmath}                                                                                          
\usepackage{graphicx}%                                                                                                      

\usepackage{epsf}
\usepackage{epsfig}
\usepackage{epstopdf}
\usepackage{hyperref}
\usepackage{amsmath}
\usepackage{amsfonts,amssymb}

\usepackage{colordvi}
\usepackage{color}
\usepackage{lscape}
\usepackage{rotfloat}
\usepackage{rotating}
\usepackage{slashed}
\usepackage{float}
\usepackage{array}

\usepackage{dsfont}

\usepackage{booktabs}
\usepackage [english]{babel}
\newcommand{\be}{\begin{equation}}
\newcommand{\ee}{\end{equation}}
\newcommand{\bea}{\begin{eqnarray}}
\newcommand{\eea}{\end{eqnarray}}

\newcommand{\MSbar}{{\overline{\rm MS}}}

\newcommand{\pa}{\partial}

\newcommand{\la}{\lambda}

\newcolumntype{M}[1]{>{\centering\arraybackslash}m{#1}}
\newcolumntype{N}{@{}m{0pt}@{}}

\def\lsim{\mathrel{\rlap{\lower4pt\hbox{\hskip1pt$\sim$}}
    \raise1pt\hbox{$<$}}}                % less than or approx. symbol
\def\slashed{{/}\mskip-10.0mu}

\begin{document}
\vspace*{1.75cm}

\title{Gauge-invariant Renormalization of the Gluino-Glue operator}

\author{M.~Costa}\thanks{kosta.marios@ucy.ac.cy}
\affiliation{Department of Physics, University of Cyprus, Nicosia, CY-1678, Cyprus}
\affiliation{Department of Mechanical Engineering and Materials Science and Engineering, Cyprus University of Technology, Limassol, CY-3036, Cyprus}

\author{G.~Panagopoulos}\thanks{gpanago@stanford.edu}
\affiliation{Department of Physics, Stanford University, CA 94305–2004, USA}

\author{H.~Panagopoulos}\thanks{haris@ucy.ac.cy}
\affiliation{Department of Physics, University of Cyprus, Nicosia, CY-1678, Cyprus}

\author{G.~Spanoudes}\thanks{spanoudes.gregoris@ucy.ac.cy}
\affiliation{Department of Physics, University of Cyprus, Nicosia, CY-1678, Cyprus}

\begin{abstract}
We study the Gluino-Glue operator in the context of Supersymmetric ${\cal N}{=}1$ Yang-Mills (SYM) theory. This composite operator is gauge invariant, and it is directly connected to light bound states of the theory; its renormalization is very important as a necessary step for the study of low-lying bound states via numerical simulations. We make use of a Gauge-Invariant Renormalization Scheme (GIRS). This requires the calculation of the Green's function of a product of two Gluino-Glue operators, situated at distinct space-time points. Within this scheme, the mixing with non-gauge invariant operators which have the same quantum numbers is inconsequential. We compute the one-loop conversion factor relating the GIRS scheme to $\MSbar$. This conversion factor can be used in order to convert to $\MSbar$ Green's functions which are obtained via lattice simulations and are renormalized nonperturbatively in GIRS.
\end{abstract}

\maketitle

\section{Introduction}
\label{Introduction}
\bigskip

The study of Supersymmetric models on the lattice has been very limited to date, due to their sheer complexity (see e.g.~\cite{ Giedt:2009yd, Bergner:2016sbv}, and references therein). The fact that SUSY is broken explicitly on the lattice poses severe problems to its correct simulation and to the numerical study of spontaneous SUSY breaking. A thorough renormalization procedure is an essential prerequisite towards non-perturbative investigations~\cite{Giedt:2004qs}. This procedure must determine all relevant Lagrangian counterterms, and all renormalizations and mixing coefficients of operators, so that the correct continuum limit can be reached, with SUSY and chiral symmetry restored in this limit. A most appropriate prototype theory, exhibiting all the above features and including both gauge and matter fields, is Supersymmetric Quantum Chromodynamics (SQCD). The study of SQCD is already very complicated due to its many degrees of freedom and interaction terms~\cite{Giedt:2009yd, Costa:2017rht, Costa:2018mvb}. Consequently, the study of composite operators and their mixing is presently out of reach, especially at the nonperturbative level. A simpler theory, and an important forerunner to the more complex models, is the Supersymmetric Yang-Mills theory (SYM)~\cite{Curci:1986sm, Catterall:2014vga, Ali:2018fbq}. It contains only gauge fields and it exhibits an interesting spectrum of bound states, in particular particles made of gluino ($\lambda$) and gluon $(u_\mu)$ fields. Preliminary nonperturbative investigations in this direction were performed in Refs.~\cite{Ali:2019agk, Ali:2018dnd, Ali:2020mvj}. 
A fundamental ingredient in these investigations is the Gluino-Glue composite operator, ${\cal O}_{Gg}$ (see Eq.~(\ref{GgO}) for its definition). It is the gauge invariant operator of lowest possible dimensionality, containg both the gluino and gluon fields.

 An important issue, which needs to be addressed in order to obtain meaningful results from lattice investigations, is the renormalization of ${\cal O}_{Gg}$ in a nonperturbative manner. Perturbation theory plays a crucial role in the development of a complete renormalization prescription, which deals with divergences and possible mixing. In a previous study~\cite{Costa:2020keq}, we identified the set of operators which mix with ${\cal O}_{Gg}$ in lattice SYM, and we calculated the one-loop renormalization factor and mixing coefficients for converting bare lattice results to the $\MSbar$ scheme, which is the standard scheme used in the analysis of experimental data. From that work, the conversion factors between an intermediate scheme, such as an RI$'$-like\footnote{RI$'$ = regularization independent}~\cite{Bochicchio:1985xa, Martinelli:1994ty}, which is directly applicable in lattice simulations, and $\MSbar$ can be easily extracted, leading to the determination of a nonperturbative renormalization prescription. However, in such a scheme the calculation of Green's functions of operators, which are gauge noninvariant and/or depend on ghost fields, is unavoidable. Non perturbative studies of ghost terms by lattice simulations are nontrivial; similarly, evaluation of gauge-dependent Green's functions on the lattice is complicated by the presence of Gribov copies. Thus, the RI$'$ scheme involves a complex mixing pattern in perturbation theory, and it is further afflicted with conceptual issues on the lattice. Therefore, a gauge invariant renormalization scheme represents a very good alternative in order to avoid these issues.

In a recent study~\cite{Costa:2021}, we introduced a gauge-invariant renormalization scheme (GIRS), in the spirit of the coordinate space (X-space) scheme~\cite{Gimenez:2004me, Gracey:2009da, Cichy:2012is, Tomii:2018zix}, which is suitable for renormalizing gauge-invariant operators nonperturbatively, without the issues of ghost terms and gauge fixing. This scheme involves Green's functions of two or more composite operators in different spacetime points, without external elementary fields (see, e.g., Eq.~(\ref{GFsGg})). Gauge-noninvariant operators do not contribute in such Green's functions, and thus, they need not be considered any further. As a consequence, the set of mixing operators in GIRS is greatly reduced and in particular, includes only gauge-invariant operators, which are accessible by lattice simulations. In the case of ${\cal O}_{\rm Gg}$, there are no gauge-invariant operators which mix with ${\cal O}_{Gg}$; thus, by applying GIRS we only need to calculate one Green's function (Eq.~(\ref{GFsGg})) to extract the multiplicative renormalization factor of ${\cal O}_{Gg}$, which contributes in physical matrix elements.

In this study, we employ GIRS in the renormalization of the Gluino-Glue operator and we present our one-loop results for the conversion factor between GIRS and $\MSbar$. This is the first time that GIRS is applied to lattice studies of Supersymmetry. This is a promising prescription, especially for nonperturbative investigations in lattice simulations, since systematic errors due to operator mixing are eliminated. Also, as described in Ref.~\cite{Costa:2021}, appropriate variants of GIRS may lead to reduced statistical noise in Monte Carlo simulations. A good candidate in this respect involves the integration over timeslices of the insertion point of an operator in a Green's function (``t-GIRS'').% is  for yielding more reliable results. %For example, the integration of the points of operator insertion in Green's functions over timeslices is a good candidate for giving more reliable results.

The paper is organized as follows: Sec.~\ref{sec2} contains all relevant definitions and the calculational setup. We also provide results on the bare Green's function which enters the GIRS procedure and we confirm the $\MSbar$ renormalization factor of ${\cal O}_{Gg}$. In Sec.~\ref{sec3} we present the GIRS renormalization prescription, along with its variant t-GIRS, and we compute the conversion factors between these schemes and $\MSbar$. Finally, we conclude in Sec.~\ref{summary} with a summary and a discussion of our results as well as of possible future extensions of our work.

\section{Formulation and $\MSbar$ calculation}
\label{sec2}
In this Section we introduce the setup for our calculation. Since we are ultimately interested in converting nonperturbatively renormalized quantities to the $\MSbar$ scheme, a necessary prerequisite is the perturbative evaluation of these quantities in dimensional regularization (DR), followed by renormalization in $\MSbar$.

Our work focuses on the supersymmetric Yang-Mills theory with gauge group $SU(N_c)$ and ${\cal N}=1$ supersymmetry generators. In the Wess-Zumino gauge~\cite{Wess:1992cp}, in which several components of the initial vector superfield vanish, the Lagrangian contains a gluon field ($u_\mu$), a gluino field ($\lambda_M$) and an auxiliary field ($D$); in standard notation it reads:
%The action of SYM in Minkowski space is ($D^{\alpha}$ is an auxiliary field):
\begin{equation} \label{SYMD}
\mathcal{L}_{SYM}=-\frac{1}{4}u_{\mu \nu}^{\alpha}u_{\mu \nu}^{\alpha}+\frac{i}{2}\bar{\lambda}^{\alpha}_{M}\gamma^{\mu}\mathcal{D}_{\mu}\lambda^{\alpha}_{M}+\frac{1}{2}D^{\alpha}D^{\alpha},  \quad\quad \la_M = \left( {\begin{array}{c} \la_a\\ \bar \la^{\dot a} \end{array} } \right),
\end{equation}
[$\alpha, \beta, \gamma = 1,\ldots,N_c^2-1$ are color indices in the adjoint representation; $a, \dot a = 1, 2$ are spinorial indices]. The subscript $M$ recalls the Majorana nature of the gluino. Henceforth we will omit this subscript for simplicity. The field strength $u_{\mu \nu}$ and the covariant derivarive of $\la$ are:
\bea
u_{\mu \nu}^{\alpha}&=&\partial_{\mu}u_{\nu}^{\alpha}-\partial_{\nu}u_{\mu}^{\alpha}-g f^{\alpha \beta \gamma}u_{\mu}^{\beta}u_{\nu}^{\gamma},\nonumber\\
\mathcal{D}_{\mu}\lambda^{\alpha}&=&\partial_{\mu}\lambda^{\alpha}-g f^{\alpha \beta \gamma}u_{\mu}^{\beta}\lambda^{\gamma},
\eea
[$f^{\alpha \beta \gamma}$ are the structure constants of the $su(N_c)$ algebra].

By eliminating the auxiliary field, we get:
\begin{equation} \label{SYMEq1}
\mathcal{L}_{SYM}=-\frac{1}{4}u_{\mu \nu}^{\alpha}u_{\mu \nu}^{\alpha}+\frac{i}{2}\bar{\lambda}^{\alpha}\gamma^{\mu}\mathcal{D}_{\mu}\lambda^{\alpha}.
\end{equation}
${\cal L}_{\rm SYM}$ is invariant up to a total derivative under the supersymmetry transformations with Grassmann spinor parameter $\xi$:
\bea
\delta_\xi u_\mu^{\alpha} & = & -i \bar \xi \gamma^\mu \lambda^{\alpha}, \nonumber \\
\delta_\xi \lambda^{\alpha} & = & \frac{1}{4} u_{\mu \nu}^{\alpha} [\gamma^{\mu},\gamma^{\nu}] \xi \,.
\label{susytransfDirac}
\eea

Gauge trasformations act on the fields as:
\begin{align}
u'_\mu &= G^{-1} u_\mu G + \frac{i}{g} (\partial_\mu G^{-1})G, &\lambda' &= G^{-1} \lambda G, &
\label{SgaugeTranComponents}
\end{align}
where $G(x) \equiv e^{i \omega^{\alpha}(x) T^{\alpha}}$, $T^\alpha$ are the generators of $su(N_c)$  and $\omega^\alpha(x)$ are real parameters.
Given that physical observables cannot depend on the choice of a gauge fixing term, and given that many regularizations, in particular the lattice regularization, violate supersymmetry at intermediate steps, one may as well choose the standard covariant gauge fixing term, proportional to $(\partial_\mu u^\mu)^2$, rather than a supersymmetric variant~\cite{Miller:1983pg},~\cite{Costa:2017rht}. The full SYM action thus includes a gauge-fixing term and a ghost term arising from the Faddeev-Popov procedure:
\begin{equation}
{\cal S}_{GF}= -\frac{1}{2\alpha}\,\int d^4x \, \left(\partial^\mu u_\mu^\alpha\right)^2,
\label{sgf}
\end{equation}
where $\alpha$ is the gauge parameter [$\alpha=1(0)$ corresponds to Feynman (Landau) gauge], and 
\begin{equation}
{\cal S}_{Ghost}= - \int d^4x \, \left( \bar{c}^\alpha\, \partial^{\mu}D_\mu^{\alpha \beta}  c^\beta\right).
\label{sghost}
\end{equation}
The ghost field $c$ is a Grassmann scalar which transforms in the adjoint representation of the gauge group, and ${\cal{D}}_\mu c =  \pa_{\mu} c + i g \,[u_\mu,c]$. Consequently, the total action in the continuum has the form:
\begin{equation}
{\cal S}_{\rm total SYM} = {\cal S}_{\rm SYM} + {\cal S}_{GF} + {\cal S}_{Ghost}.
\label{ScontALL}
\end{equation}
By construction, ${\cal S}_{\rm total SYM}$ is not gauge invariant; however it is invariant under Becchi-Rouet-Stora-Tyutin (BRST) transformations. The latter involve parameters that take their values in a Grassmann algebra. The BRST trasformations for the fields of the full SYM action can be found by setting $\omega^{\alpha}$ in Eq.~(\ref{SgaugeTranComponents}) equal to $c^{\alpha} \eta$, where $\eta$  is a Grassmann variable. 

In standard notation, the Gluino-Glue operator, ${\cal O}_{Gg}$, is defined as\footnote{For ease of notation, we leave out the free Dirac index in ${\cal O}_{Gg}$. Lorentz and color indices are saturated. Similarly, the Green's function, $G(x-y)$, has two implicit Dirac indices.}:
\be
{\cal O}_{Gg} = \sigma_{\mu \nu} \,{\rm{tr}}_c (\, u_{\mu \nu} \lambda ), \quad \sigma_{\mu \nu}=\frac{1}{2} [\gamma_{\mu},\gamma_{\nu}].
\label{GgO}
\ee

Acting on the vacuum, ${\cal O}_{Gg}$ is expected to excite a light bound state of the theory, which is a potential supersymmetric partner of the glueballs and the gluinoballs~\cite{Veneziano:1982ah,Steinhauser:2017xqc}. 

Consider the following Green's function, containing a product of Gluino-Glue operators, whose 4-vector positions $x$ and $y$ are distinct:
\be
G(x-y) \equiv \langle {\cal O}_{Gg} (x) {\overline{\cal O}}_{Gg} (y) \rangle. 
\label{GFsGg}
\ee
In order to contract gluino fields in Feynman diagrams it is convenient to choose the charge conjugate operator ${\overline{\cal O}}_{Gg}(y)$, instead of ${\cal O}_{Gg}(y)$, as the second factor in Eq.~(\ref{GFsGg});  ${\overline{\cal O}}_{Gg}(y)$ is defined as as (cf. Eq.~(\ref{GgO})):
\be
{\overline{\cal O}}_{Gg} =  -\,{\rm{tr}}_c (\, \bar \lambda u_{\mu \nu} ) \sigma_{\mu \nu}.%, \quad \sigma_{\mu \nu}=\frac{1}{2} [\gamma_{\mu},\gamma_{\nu}]
%\label{GgO}
\ee
Disconnected Feynman diagrams are not present in $G(x-y)$ due to the fact that ${\cal O}_{Gg}$ is not a scalar operator. In addition, no lower dimensional operator mixes with ${\cal O}_{Gg}$. In the following, we first calculate the tree-level Green's function, where we regularize the theory in $d$ dimensions ($d=4-2\,\epsilon$). Given that ${\cal O}_{Gg}$ has a relatively high dimensionality, 7/2, there is a number of other (non-gauge invariant!) composite operators with which ${\cal O}_{Gg}$ can, and will, mix~\cite{Costa:2020keq} ; this becomes apparent when one calculates Green's functions with elementary external fields, as is done in a typical renormalization procedure. A proper treatment of this mixing entails studying the 2-point and 3-point Green's functions of ${\cal O}_{Gg}$ with external gluino and gluon fields. On the other hand, if one uses the GIRS scheme all non-gauge invariant operators will not contribute to $G(x-y)$, and one will obtain directly the multiplicative renormalization of ${\cal O}_{Gg}$, which is the only renormalization factor which is relevant for physical matrix elements.

\begin{figure}[H]
\centering
\epsfig{file=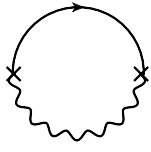,scale=0.334}
\caption{Tree-level Feynman diagram contributing to the expectation value $\langle {\cal O}_{Gg}(x) {\cal O}_{Gg}(y)\rangle$. A wavy (solid) line represents gluons (gluinos). A cross denotes the insertion of the Gluino-Glue operator.}
\label{DRtree}
\end{figure}

The Feynman diagram for the tree-level value of the 2-point Green's function is shown in Fig.~\ref{DRtree}. 

The tree-level contribution has the following momentum-integral form:
\be
G(x-y)^{\rm tree} = -4 i (N_c^2 - 1) (d-2) \int \frac{d^d p}{(2 \pi)^d} \int \frac{d^d k}{(2 \pi)^d} \ e^{-i (x-y) \cdot k} \frac{\slashed{p} \ (p \cdot k)}{p^2 (p-k)^2}.
\ee
Integrating over the loop momenta, and after some manipulations of Dirac matrices, the resulting expression is:
\begin{equation}
%\langle {\cal O}_{Gg} (x) {\cal O}_{Gg} (y) \rangle^{{\rm tree}} = 
G(x-y)^{\rm tree} = -2\, \frac{(N_c^2-1)\:\Gamma(2-\epsilon)^2}{\pi^{4-2\epsilon}} (-1 + \epsilon)(-3 + 2\epsilon) \,\slashed{z}\,(z^2)^{-4 + 2\epsilon}, \quad z^\mu \equiv y^\mu - x^\mu.
 \label{Gtree1}
\end{equation}

\begin{figure}[H]
\centering
\epsfig{file=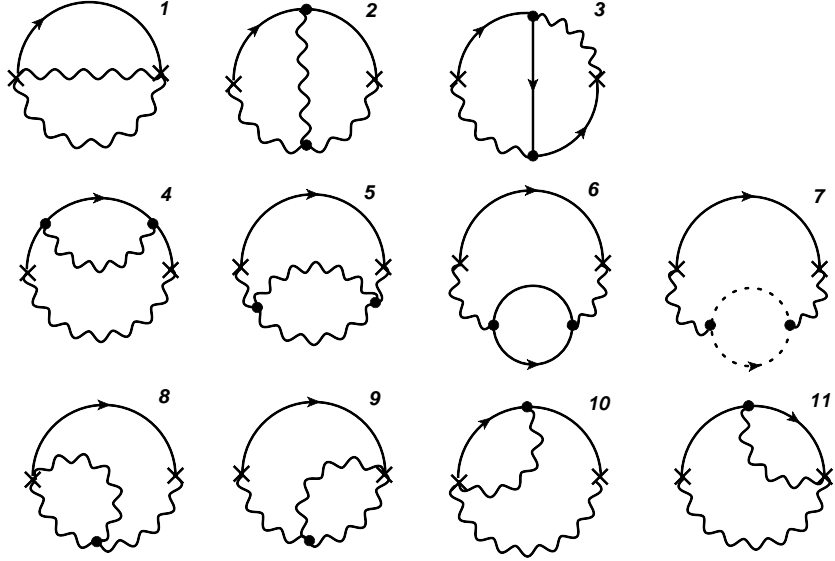,scale=0.667}
\caption{One-loop Feynman diagrams contributing to the expectation value $\langle {\cal O}_{Gg}(x) {\cal O}_{Gg}(y)\rangle$. A wavy (solid) line represents gluons (gluinos). The dashed line is the ghost field. A cross denotes the insertion of the operator. }
\label{oneloop}
\end{figure}

Figure~\ref{oneloop} shows the Feynman diagrams contributing to $G(x-y)$ at one loop.

By rotational invariance, the expression for each diagram is proportional to the tree-level Green's function. Diagrams 3, 6, 7 are independent of $\alpha$, while $\alpha^1$ terms occur in diagrams 1-2, 4-5, 8-11, and $\alpha^2$ terms occur in diagrams 1, 5, 8-9. Upon summation over all diagrams, $\alpha$ terms vanish and thus the resulting expression is gauge independent. 

By adding tree-level and one-loop contributions, the bare Green's function takes the following form:
\bea
%  \langle \mathcal{O}_{Gg} (x) \mathcal{O}_{Gg} (y) \rangle^{\rm bare} 
G(x-y)^{\rm bare} &{=}& 
%\langle \mathcal{O}_{Gg} (x) \mathcal{O}_{Gg} (y) \rangle^{\rm tree} 
G(x-y)^{\rm tree} \times \nonumber \\
 && \Bigg\{ 1 - \frac{g^2 \ N_c}{16 \pi^2} \left(\bar{\mu}^2 z^2 \right)^\epsilon \frac{e^{\epsilon \gamma_E} \Gamma(-\epsilon)} {4^\epsilon \ \epsilon \ (1 - \epsilon)^3 \ (3 - 2\epsilon)} \times \nonumber \\
 && \qquad \qquad \quad \ \Bigg[ (1-\epsilon) \ (12 -48 \epsilon + 70 \epsilon^2 - 39 \epsilon^3 + \epsilon^4) + \nonumber \\
&&  \qquad \qquad \qquad \frac{(1 - 3 \epsilon + 2 \epsilon^2 + \epsilon^3) \
    \Gamma(-\epsilon) \ \Gamma(\epsilon)^2 \ \Gamma(4-3\epsilon)}{4 \ (1-2\epsilon) \ \Gamma(-2\epsilon)^2 \ \Gamma(2\epsilon)} \Bigg] + \mathcal{O} (g^4) \Bigg\},
 \label{GFtreePoneloop}
\eea
where $\bar\mu$ is the $\MSbar$ renormalization scale 4-vector relating the dimensionful coupling constant $g_L$ in the $d$-dimensional Lagrangian to the dimensionless ``bare'' coupling constant $g^B$: $g_L = \mu^\epsilon g^B$ ($\mu = \bar \mu \sqrt{e^{\gamma_E}/ 4\pi}$). To this perturbative order, the distinction between bare and renormalized coupling constants is inessential; we will thus denote both by $g$. 

Evaluation of $G(x-y)$ to order $g^{2n}$ involves diagrams with $(n+1)$ loops. This is a nonnegligible price to pay. However, all these diagrams involve massless fields and (upon expressing them in momentum space) only one incoming/outgoing momentum; such diagrams can be evaluated to very high perturbative order (see, e.g., \cite{Chetyrkin:2010dx, Baikov:2014qja,  Luthe:2017ttc, Chetyrkin:2017bjc, Herzog:2018kwj} for up to 5-loop calculations). 

The renormalization factor $Z_{Gg}^{B,R}$ relating the bare Gluino-Glue operator in the ``B'' regularization to the renormalized operator in the ``R'' renomalization scheme is defined by:
\be
\mathcal{O}_{Gg}^R = Z_{Gg}^{B,R} \mathcal{O}_{Gg}^B + {\rm other \, operators \, which \, will \, not\, contribute \,in}\,\, G(x-y).
\ee
In (DR, $\MSbar$), the renormalization factor $Z_{Gg}^{{\rm DR},\MSbar}$ is defined to have only negative integer powers of $\epsilon$ and the following condition is imposed:
\be
\left[\left(Z_{Gg}^{{\rm DR},\MSbar}\right)^2 \langle \mathcal{O}_{Gg} (x) \mathcal{O}_{Gg} (y) \rangle^{\rm bare} \right]\Big|_{\epsilon^{-n}} = 0, \quad n \in \mathbb{Z}^+.
\label{condMSbar}
\ee 
%Notice that the expression for the renormalization of the Gluino-Glue operator in the $\MSbar$ renormalization scheme is also gauge independent; this was expected given that the renormalization condition relies on a gauge-invariant Green's function. 

By removing the pole parts in the bare Green's function one defines the $\MSbar$-renormalized Green's function and determines the renormalization factor $Z_{Gg}^{DR,\MSbar}$. Thus, the $\MSbar$-renormalized 2-point Green's function takes the following form:
\be
  \langle \mathcal{O}_{Gg}^{\MSbar} (x) \mathcal{O}_{Gg}^{\MSbar} (y) \rangle = \frac{6 \ (N_c^2 - 1) \ \slashed{z}}{\pi^4 (z^2)^4} \ \Bigg\{ 1 + \frac{g^2}{16 \pi^2} \Bigg[ \frac{2 N_c}{3} \left(5 + 18 \gamma_E - 9 \ln(4) + 9 \ln(\bar{\mu}^2 z^2)\right) \Bigg] + \mathcal{O} (g^4) \Bigg\}.
 \label{GFtreePoneloopMSbar}
\ee
The extraction of the correct renormalization in the $\MSbar$ scheme, already known in Ref.~\cite{Costa:2020keq} for the operator ${\cal O}_{Gg}$ is a consistency check of our results. From the renormalization condition Eq~(\ref{condMSbar}), at one-loop order, we find:
\be
Z_{Gg}^{DR,\MSbar} = 1 - \frac{g^2 N_c}{16\pi^2} \frac{3}{\epsilon}.
\ee

\section{Renormalization factor of the Gluino-Glue operator in the GIRS scheme}
\label{sec3}

The purpose of this section is to find the conversion factor between $\MSbar$ renormalization and the GIRS renormalization of ${\cal O}_{Gg}$. The renormalization factor in GIRS can be obtained by imposing the following condition on the renormalized Green's function: 

\begin{eqnarray}
{\rm Tr} \Big[ (\slashed{y} - \slashed{x}) \,\langle {\mathcal O}^{\rm GIRS}_{Gg} (x)
{{\mathcal O}^{\rm GIRS}_{Gg}} (y) \rangle \Big]
\vert_{y-x = \bar{z}} & \equiv & 
(Z^{B,{\rm GIRS}}_{Gg})^2 \ {\rm Tr} \Big[(\slashed{y} - \slashed{x}) \, \langle {\mathcal O}^B_{Gg} (x)
{{\mathcal O}^B_{Gg}} (y) \rangle \Big]
\vert_{y-x = \bar{z}} \nonumber \\
&=& {\rm Tr} \Big[ (\slashed{y} - \slashed{x})\,\langle {\mathcal O}^{B}_{Gg} (x)
{{\mathcal O}^{B}_{Gg}} (y) \rangle^{\rm tree} \Big] \vert_{y-x = \bar{z}}, \label{GIRScond}   
\end{eqnarray}
where the 4-vector $\bar{z}$ is the GIRS renormalization scale ($\bar{z} \neq 0$). In the middle expression of Eq.~(\ref{GIRScond}) it is implicit that the regulator is sent to its limiting value ($\epsilon \to 0$ in DR). As we are interested in applying GIRS in lattice simulations, the scale $\bar{z}$ may be chosen to satisfy the condition $a \ll |\bar{z}| \ll \Lambda_{\rm SYM}^{-1}$, where $a$ is the lattice spacing and $\Lambda_{\rm SYM}$ is the SYM physical scale; this condition guarantees that discretization effects will be under control and simultaneously we will be able to make contact with (continuum) perturbation theory. 

The ratio between the $\MSbar$ and ${\rm GIRS}$ renormalization factors gives the corresponding conversion factor:
\be
C_{Gg}^{\MSbar,{\rm GIRS}} =Z_{Gg}^{DR,\MSbar}/Z_{Gg}^{DR,{\rm GIRS}}.
\ee
Being regularization independent, the same conversion factor can then be also used in the lattice regularization (L): 
\be
C_{Gg}^{\MSbar,{\rm GIRS}} = Z_{Gg}^{L,\MSbar}/ Z_{Gg}^{L,{\rm GIRS}}.
\ee
Combining the {\it ipso facto} perturbative evaluation of $C_{Gg}^{\MSbar,GIRS}$ with the nonperturbative evaluation of $Z_{Gg}^{L,GIRS}$, one is thus led to the desired renormalization factor $Z_{Gg}^{L,\MSbar}$.

Combining Eqs.~(\ref{GFtreePoneloopMSbar}), (\ref{GIRScond}) with the relation:
\be
{\rm Tr} \Big[ (\slashed{y} - \slashed{x}) \,\langle {\mathcal O}^{\MSbar}_{Gg} (x) {{\mathcal O}^{\MSbar}_{Gg}} (y) \rangle \Big]
\vert_{y-x = \bar{z}} \equiv  (C_{Gg}^{\MSbar,{\rm GIRS}})^2 \ {\rm Tr} \Big[(\slashed{y} - \slashed{x}) \, \langle {\mathcal O}^{\rm GIRS}_{Gg} (x) {{\mathcal O}^{\rm GIRS}_{Gg}} (y) \rangle \Big] \vert_{y-x = \bar{z}},
\label{CGgMStoGIRS}
\ee
we obtain the result for the conversion factor:

\be
 C_{Gg}^{\MSbar, {\rm GIRS}} = 1 + \frac{g_\MSbar^2 N_c}{16 \pi^2} \Big(\frac{5}{3}
 + 6 \gamma_E - 3 \ln (4) + 3 \ln (\bar{\mu}^2 \bar{z}^2) \Big)
 + \mathcal{O} (g^4).
\ee

There are additional, alternative ways for extracting renormalization factors in GIRS, using variants of the Green's functions of Eq. (\ref{GFsGg}). An option is to take Fourier transform of Eq.(\ref{GFsGg}); however, this is not an optimal choice as contact terms arise. A more promising option is to integrate Eq.(\ref{GFsGg}) over three of the four components of the position vector $(x - y)$, while setting the fourth component equal to a reference scale $\bar{t}$. 
Due to the anisotropic lattice employed in simulations, the temporal direction is a special one. In this sense, a natural choice for the component $\bar{t}$ is to be temporal; we call this variant ``t-GIRS''. Without loss of generality, we set $x=(x_1,x_2,x_3,0)$ and $y=(0,0,0,\bar{t})$; then the renormalization condition for t-GIRS has the following form:
%\begin{equation}
%\int d^3\vec{x} \ \langle O^{\rm t-GIRS}_{Gg} (\vec{x}, 0) {O^{\rm t-GIRS}_{Gg}} (\vec{0},\bar{t}) \rangle = \int d^3\vec{x} \ \langle O^{\rm t-GIRS}_{Gg} (\vec{x}, 0) {O^{\rm t-GIRS}_{Gg}} (\vec{0}, \bar{t}) \rangle^{\rm tree}. \label{RC2}
%\end{equation}  
\begin{equation}
{\rm Tr} \Big[ \Big( \int d^3\vec{x} \ \langle {\mathcal O}^{\rm t-GIRS}_{Gg} (\vec{x}, 0) {{\mathcal O}^{\rm t-GIRS}_{Gg}} (\vec{0},\bar{t}) \rangle \Big) \gamma_4 \Big] = {\rm Tr} \Big[ \Big( \int d^3\vec{x} \ \langle {\mathcal O}^B_{Gg} (\vec{x}, 0) {{\mathcal O}^B_{Gg}} (\vec{0}, \bar{t}) \rangle^{\rm tree} \Big) \gamma_4 \Big],
\label{t-GIRS_condition}
\end{equation}
where the tree-level Green's function in the right-hand side of the above condition is given by Eq.~\eqref{Gtree1} with $\epsilon \rightarrow 0$. By analogy with Eq.~\ref{CGgMStoGIRS}, we obtain the conversion factor between $\MSbar$ and t-GIRS from the relation:
% For extracting the renormalization factor of $\mathcal{O}_{Gg}$ in DR, one express the above condition in terms of the $\MSbar$-renormalized Green's function, which is regulator-independent, since the integration over the spatial components is taken in 4, rather than D, dimensions:
\begin{equation}
{\rm Tr} \Big[ \Big( \int d^3\vec{x} \ \langle {\mathcal O}^{\MSbar}_{Gg} (\vec{x}, 0) {{\mathcal O}^{\MSbar}_{Gg}} (\vec{0},\bar{t}) \rangle \Big) \gamma_4 \Big] = \left(C_{Gg}^{\MSbar,{\rm t-GIRS}}\right)^2 {\rm Tr} \Big[ \Big( \int d^3\vec{x} \ \langle {\mathcal O}^B_{Gg} (\vec{x}, 0) {{\mathcal O}^B_{Gg}} (\vec{0}, \bar{t}) \rangle^{\rm tree} \Big) \gamma_4 \Big].
\label{t-GIRS_condition2}
\end{equation}
Note that both sides of Eq.~(\ref{t-GIRS_condition2}) are well-defined as $d \to 4$, and thus spatial integration is performed in $3$ (rather than $d-1$) dimensions.

Use of the integrals:
\be
\int d^3\vec{x} \ \frac{1}{(|\vec{x}|^2 + t^2)^4} = \frac{\pi^2}{8 |t|^5}, \quad \int d^3\vec{x} \ \frac{\ln(|\vec{x}|^2 + t^2)}{(|\vec{x}|^2 + t^2)^4} = \frac{\pi^2 (-5 + 12 \ln(2) + 6 \ln (t^2))}{48 |t|^5},
\ee
in Eq.~(\ref{t-GIRS_condition2}), leads to the following expression for the conversion factor from  t-GIRS to $\MSbar$:
\begin{equation}
C_{Gg}^{\MSbar, {\rm t-GIRS}} = 1 + \frac{g^2 N_c}{16 \pi^2} \Big(-\frac{5}{6} + 6 \gamma_E + 3 \ln (\bar{\mu}^2 \bar{t}^2) \Big) + \mathcal{O} (g^4).
\end{equation}
Using a standard lattice discretization ${\cal O}_{Gg}^L$ of the Gluino-Glue operator~\cite{Ali:2018dnd}, the non-perturbative value of its renormalization factor, $Z_{Gg}^{L, {\rm t-GIRS}}$, can be found via Eq.~(\ref{CGgMStoGIRS}) as follows:
\be
(Z_{Gg}^{L,{\rm t-GIRS}})^2 {\rm Tr} \Big[ \Big( \int d^3\vec{x} \ \langle {\mathcal O}^{L}_{Gg} (\vec{x}, 0) {{\mathcal O}^{L}_{Gg}} (\vec{0},\bar{t}) \rangle \Big) \gamma_4 \Big] = 
{\rm Tr} \Big[ \Big( \int d^3\vec{x} \, \lim_{\epsilon \to 0}\langle {\cal O}_{Gg} (x) {\cal O}_{Gg} (y) \rangle^{{\rm tree}}  \Big) \gamma_4 \Big] = \frac{3}{4} \frac{(N_c^2-1)}{\pi^{2} t^{4}}.
 \label{Gtree4dim}
\ee

\section{Summary of Results and future plans}
\label{summary}

In this paper we have studied the Gluino-Glue operator ${\cal O}_{Gg}$ in SYM theory, using variants of a gauge invariant renormalization scheme (GIRS).
In this scheme, renormalization factors for physical observables are defined via correlation functions of gauge invariant operators (in coordinate space, so as to avoid unwanted contact terms). The advantage in doing so is twofold: On one hand, one may safely ignore the existence of mixing with gauge variant operators, possibly containing ghost fields, as is the case with ${\cal O}_{Gg}$\,; on the other hand, all necessary correlation functions can be computed nonperturbatively in numerical simulations, without need for gauge fixing or treating ghost fields. In order to achieve renormalization in a more standard scheme, such as $\MSbar$, one must calculate appropriate conversion factors; these must necessarily rely on perturbation theory, by the very definition of $\MSbar$, and, being regularization independent, can be most naturally computed in dimensional regularization. The downside of this scheme is that the extraction of conversion factors to order $g^{2n}$ requires evaluation of $(n+1)$-loop Feynman diagrams; however, such diagrams are typically two-point and massless, and thus their evaluation can be carried out to very high perturbative order.

We have evaluated the conversion factor for ${\cal O}_{Gg}$ to order $g^2$, using two variants of GIRS, by calculating the two-loop diagrams in the correlation function involving a product of two ${\cal O}_{Gg}$ operators at distinct positions. Some natural extensions of our work, besides going to higher perturbative order, are: (i) Addressing SQCD, where the inclusion of quarks and squarks causes ${\cal O}_{Gg}$ to mix with other gauge invariant operators, even (on the lattice) one of lower dimensionality, (ii) Study of other fermionic operators, notably the Noether supercurrent, which is instrumental in nonperturbative studies of Ward identities and restoration of SUSY on the lattice. The supercurrent presents operator mixing as well, and its renormalization in GIRS can be performed along the lines of Ref.~\cite{Costa:2021}.

\begin{acknowledgements}
M.C. and H.P. acknowledge financial support from the project \textit{``Quantum Fields on the Lattice''}, funded by the Cyprus Research and Innovation Foundation (RIF) under contract number EXCELLENCE/0918/0066. G.S. acknowledges financial support by the University of Cyprus, under the research programs entitled \textit{``Quantum Fields on the Lattice''} and \textit{``Nucleon parton distribution functions using Lattice Quantum Chromodynamics''}. 
\end{acknowledgements}

\bibliographystyle{elsarticle-num}                     % Style for bibliography
\bibliography{GIRS_Gg_References}

\begin{thebibliography}{10}
\expandafter\ifx\csname url\endcsname\relax
  \def\url#1{\texttt{#1}}\fi
\expandafter\ifx\csname urlprefix\endcsname\relax\def\urlprefix{URL }\fi
\expandafter\ifx\csname href\endcsname\relax
  \def\href#1#2{#2} \def\path#1{#1}\fi

\bibitem{Giedt:2009yd}
J.~Giedt, {Progress in four-dimensional lattice supersymmetry}, Int. J. Mod.
  Phys. A 24 (2009) 4045--4095.
\newblock \href {http://arxiv.org/abs/0903.2443} {\path{arXiv:0903.2443}},
  \href {http://dx.doi.org/10.1142/S0217751X09045492}
  {\path{doi:10.1142/S0217751X09045492}}.

\bibitem{Bergner:2016sbv}
G.~Bergner, S.~Catterall, {Supersymmetry on the lattice}, Int. J. Mod. Phys. A
  31~(22) (2016) 1643005.
\newblock \href {http://arxiv.org/abs/1603.04478} {\path{arXiv:1603.04478}},
  \href {http://dx.doi.org/10.1142/S0217751X16430053}
  {\path{doi:10.1142/S0217751X16430053}}.

\bibitem{Giedt:2004qs}
J.~Giedt, E.~Poppitz, {Lattice supersymmetry, superfields and renormalization},
  JHEP 09 (2004) 029.
\newblock \href {http://arxiv.org/abs/hep-th/0407135}
  {\path{arXiv:hep-th/0407135}}, \href
  {http://dx.doi.org/10.1088/1126-6708/2004/09/029}
  {\path{doi:10.1088/1126-6708/2004/09/029}}.

\bibitem{Costa:2017rht}
M.~Costa, H.~Panagopoulos, {Supersymmetric QCD on the Lattice: An Exploratory
  Study}, Phys. Rev. D 96~(3) (2017) 034507.
\newblock \href {http://arxiv.org/abs/1706.05222} {\path{arXiv:1706.05222}},
  \href {http://dx.doi.org/10.1103/PhysRevD.96.034507}
  {\path{doi:10.1103/PhysRevD.96.034507}}.

\bibitem{Costa:2018mvb}
M.~Costa, H.~Panagopoulos, {Supersymmetric QCD: Renormalization and Mixing of
  Composite Operators}, Phys. Rev. D 99~(7) (2019) 074512.
\newblock \href {http://arxiv.org/abs/1812.06770} {\path{arXiv:1812.06770}},
  \href {http://dx.doi.org/10.1103/PhysRevD.99.074512}
  {\path{doi:10.1103/PhysRevD.99.074512}}.

\bibitem{Curci:1986sm}
G.~Curci, G.~Veneziano, {Supersymmetry and the Lattice: A Reconciliation?},
  Nucl. Phys. B 292 (1987) 555--572.
\newblock \href {http://dx.doi.org/10.1016/0550-3213(87)90660-2}
  {\path{doi:10.1016/0550-3213(87)90660-2}}.

\bibitem{Catterall:2014vga}
S.~Catterall, J.~Giedt, D.~Schaich, P.~H. Damgaard, T.~DeGrand, {Results from
  lattice simulations of N=4 supersymmetric Yang--Mills}, PoS LATTICE2014
  (2014) 267.
\newblock \href {http://arxiv.org/abs/1411.0166} {\path{arXiv:1411.0166}},
  \href {http://dx.doi.org/10.22323/1.214.0267}
  {\path{doi:10.22323/1.214.0267}}.

\bibitem{Ali:2018fbq}
S.~Ali, H.~Gerber, I.~Montvay, G.~M\"unster, S.~Piemonte, P.~Scior, G.~Bergner,
  {Analysis of Ward identities in supersymmetric Yang\textendash{}Mills
  theory}, Eur. Phys. J. C 78~(5) (2018) 404.
\newblock \href {http://arxiv.org/abs/1802.07067} {\path{arXiv:1802.07067}},
  \href {http://dx.doi.org/10.1140/epjc/s10052-018-5887-9}
  {\path{doi:10.1140/epjc/s10052-018-5887-9}}.

\bibitem{Ali:2019agk}
S.~Ali, G.~Bergner, H.~Gerber, I.~Montvay, G.~M\"unster, S.~Piemonte, P.~Scior,
  {Numerical results for the lightest bound states in $\mathcal{N}=1$
  supersymmetric SU(3) Yang-Mills theory}, Phys. Rev. Lett. 122~(22) (2019)
  221601.
\newblock \href {http://arxiv.org/abs/1902.11127} {\path{arXiv:1902.11127}},
  \href {http://dx.doi.org/10.1103/PhysRevLett.122.221601}
  {\path{doi:10.1103/PhysRevLett.122.221601}}.

\bibitem{Ali:2018dnd}
S.~Ali, G.~Bergner, H.~Gerber, P.~Giudice, I.~Montvay, G.~M\"unster,
  S.~Piemonte, P.~Scior, {The light bound states of $\mathcal{N}=1$
  supersymmetric SU(3) Yang-Mills theory on the lattice}, JHEP 03 (2018) 113.
\newblock \href {http://arxiv.org/abs/1801.08062} {\path{arXiv:1801.08062}},
  \href {http://dx.doi.org/10.1007/JHEP03(2018)113}
  {\path{doi:10.1007/JHEP03(2018)113}}.

\bibitem{Ali:2020mvj}
S.~Ali, G.~Bergner, H.~Gerber, I.~Montvay, G.~M\"unster, S.~Piemonte, P.~Scior,
  {Continuum extrapolation of Ward identities in ${{\mathcal {N}}=1}$
  supersymmetric SU(3) Yang\textendash{}Mills theory}, Eur. Phys. J. C 80~(6)
  (2020) 548.
\newblock \href {http://arxiv.org/abs/2003.04110} {\path{arXiv:2003.04110}},
  \href {http://dx.doi.org/10.1140/epjc/s10052-020-8113-5}
  {\path{doi:10.1140/epjc/s10052-020-8113-5}}.

\bibitem{Costa:2020keq}
M.~Costa, H.~Herodotou, P.~Philippides, H.~Panagopoulos, {Renormalization and
  Mixing of the Gluino-Glue Operator on the Lattice, arXiv:2010.02683.}

\bibitem{Bochicchio:1985xa}
M.~Bochicchio, L.~Maiani, G.~Martinelli, G.~C. Rossi, M.~Testa, {Chiral
  Symmetry on the Lattice with Wilson Fermions}, Nucl. Phys. B262 (1985) 331.
\newblock \href {http://dx.doi.org/10.1016/0550-3213(85)90290-1}
  {\path{doi:10.1016/0550-3213(85)90290-1}}.

\bibitem{Martinelli:1994ty}
G.~Martinelli, C.~Pittori, C.~T. Sachrajda, M.~Testa, A.~Vladikas, {A General
  method for nonperturbative renormalization of lattice operators}, Nucl. Phys.
  B 445 (1995) 81--108.
\newblock \href {http://arxiv.org/abs/hep-lat/9411010}
  {\path{arXiv:hep-lat/9411010}}, \href
  {http://dx.doi.org/10.1016/0550-3213(95)00126-D}
  {\path{doi:10.1016/0550-3213(95)00126-D}}.

\bibitem{Costa:2021}
M.~Costa, I.~Karpasitis, G.~Panagopoulos, H.~Panagopoulos, T.~Pafitis,
  A.~Skouroupathis, G.~Spanoudes, {Gauge-invariant Renormalization Scheme in
  QCD: Application to fermion bilinears and the energy-momentum tensor,
  arXiv:2102.00858.}

\bibitem{Gimenez:2004me}
V.~Gimenez, L.~Giusti, S.~Guerriero, V.~Lubicz, G.~Martinelli, S.~Petrarca,
  J.~Reyes, B.~Taglienti, E.~Trevigne, {Non-perturbative renormalization of
  lattice operators in coordinate space}, Phys. Lett. B598 (2004) 227--236.
\newblock \href {http://arxiv.org/abs/hep-lat/0406019}
  {\path{arXiv:hep-lat/0406019}}, \href
  {http://dx.doi.org/10.1016/j.physletb.2004.07.053}
  {\path{doi:10.1016/j.physletb.2004.07.053}}.

\bibitem{Gracey:2009da}
J.~A. Gracey, {Three loop anti-MS operator correlation functions for deep
  inelastic scattering in the chiral limit}, JHEP 04 (2009) 127.
\newblock \href {http://arxiv.org/abs/0903.4623} {\path{arXiv:0903.4623}},
  \href {http://dx.doi.org/10.1088/1126-6708/2009/04/127}
  {\path{doi:10.1088/1126-6708/2009/04/127}}.

\bibitem{Cichy:2012is}
K.~Cichy, K.~Jansen, P.~Korcyl, {Non-perturbative renormalization in coordinate
  space for $N_f=2$ maximally twisted mass fermions with tree-level Symanzik
  improved gauge action}, Nucl. Phys. B865 (2012) 268--290.
\newblock \href {http://arxiv.org/abs/1207.0628} {\path{arXiv:1207.0628}},
  \href {http://dx.doi.org/10.1016/j.nuclphysb.2012.08.006}
  {\path{doi:10.1016/j.nuclphysb.2012.08.006}}.

\bibitem{Tomii:2018zix}
M.~Tomii, N.~H. Christ, {$O(4)$-symmetric position-space renormalization of
  lattice operators}, Phys. Rev. D99~(1) (2019) 014515.
\newblock \href {http://arxiv.org/abs/1811.11238} {\path{arXiv:1811.11238}},
  \href {http://dx.doi.org/10.1103/PhysRevD.99.014515}
  {\path{doi:10.1103/PhysRevD.99.014515}}.

\bibitem{Wess:1992cp}
J.~Wess, J.~Bagger, {Supersymmetry and supergravity}, Princeton University
  Press, Princeton, NJ, USA, 1992.

\bibitem{Miller:1983pg}
R.~D. Miller, {Supersymmetric gauge fixing and the effective potential}, Phys.
  Lett. B 129 (1983) 72--76.
\newblock \href {http://dx.doi.org/10.1016/0370-2693(83)90731-1}
  {\path{doi:10.1016/0370-2693(83)90731-1}}.

\bibitem{Veneziano:1982ah}
G.~Veneziano, S.~Yankielowicz, {An Effective Lagrangian for the Pure N=1
  Supersymmetric Yang-Mills Theory}, Phys. Lett. B 113 (1982) 231.
\newblock \href {http://dx.doi.org/10.1016/0370-2693(82)90828-0}
  {\path{doi:10.1016/0370-2693(82)90828-0}}.

\bibitem{Steinhauser:2017xqc}
M.~Steinhauser, A.~Sternbeck, B.~Wellegehausen, A.~Wipf, {Spectroscopy of
  four-dimensional $\mathcal{N}=1$ supersymmetric SU(3) Yang-Mills theory}, EPJ
  Web Conf. 175 (2018) 08022.
\newblock \href {http://arxiv.org/abs/1711.05086} {\path{arXiv:1711.05086}},
  \href {http://dx.doi.org/10.1051/epjconf/201817508022}
  {\path{doi:10.1051/epjconf/201817508022}}.

\bibitem{Chetyrkin:2010dx}
K.~G. Chetyrkin, A.~Maier, {Massless correlators of vector, scalar and tensor
  currents in position space at orders $\alpha_s^3$ and $\alpha_s^4$: Explicit
  analytical results}, Nucl. Phys. B844 (2011) 266--288.
\newblock \href {http://arxiv.org/abs/1010.1145} {\path{arXiv:1010.1145}},
  \href {http://dx.doi.org/10.1016/j.nuclphysb.2010.11.007}
  {\path{doi:10.1016/j.nuclphysb.2010.11.007}}.

\bibitem{Baikov:2014qja}
P.~A. Baikov, K.~G. Chetyrkin, J.~H. K\"uhn, {Quark Mass and Field Anomalous
  Dimensions to ${\cal O}(\alpha_s^5)$}, JHEP 10 (2014) 076.
\newblock \href {http://arxiv.org/abs/1402.6611} {\path{arXiv:1402.6611}},
  \href {http://dx.doi.org/10.1007/JHEP10(2014)076}
  {\path{doi:10.1007/JHEP10(2014)076}}.

\bibitem{Luthe:2017ttc}
T.~Luthe, A.~Maier, P.~Marquard, Y.~Schroder, {Complete renormalization of QCD
  at five loops}, JHEP 03 (2017) 020.
\newblock \href {http://arxiv.org/abs/1701.07068} {\path{arXiv:1701.07068}},
  \href {http://dx.doi.org/10.1007/JHEP03(2017)020}
  {\path{doi:10.1007/JHEP03(2017)020}}.

\bibitem{Chetyrkin:2017bjc}
K.~G. Chetyrkin, G.~Falcioni, F.~Herzog, J.~A.~M. Vermaseren, {Five-loop
  renormalisation of QCD in covariant gauges}, JHEP 10 (2017) 179, [Addendum:
  JHEP 12, 006 (2017)].
\newblock \href {http://arxiv.org/abs/1709.08541} {\path{arXiv:1709.08541}},
  \href {http://dx.doi.org/10.1007/JHEP10(2017)179}
  {\path{doi:10.1007/JHEP10(2017)179}}.

\bibitem{Herzog:2018kwj}
F.~Herzog, S.~Moch, B.~Ruijl, T.~Ueda, J.~A.~M. Vermaseren, A.~Vogt, {Five-loop
  contributions to low-N non-singlet anomalous dimensions in QCD}, Phys. Lett.
  B 790 (2019) 436--443.
\newblock \href {http://arxiv.org/abs/1812.11818} {\path{arXiv:1812.11818}},
  \href {http://dx.doi.org/10.1016/j.physletb.2019.01.060}
  {\path{doi:10.1016/j.physletb.2019.01.060}}.

\end{thebibliography}

\end{document}